\documentclass[12pt]{article}
\usepackage{amssymb}
\def\be{\begin{equation}}
\def\ee{\end{equation}}
\def\bea{\begin{eqnarray}}
\def\eea{\end{eqnarray}}
\def\p{\partial}

\def\four{\textstyle{1\over4}}

\def\pl#1{{\sl Phys.~Lett.~\bf B#1}}
\def\pr#1{{\sl Phys.~Rev.~\bf D#1}}
\def\prl#1{{\sl Phys.~Rev. Lett.~\bf #1}}

\def\cqg#1{{\sl Class.~Quant.~Grav.~\bf #1}}

\catcode`\@=11
\def\@citex[#1]#2{%
\if@filesw \immediate \write \@auxout {\string \citation {#2}}\fi
\@tempcntb\m@ne \let\@h@ld\relax \def\@citea{}%
\@cite{%
  \@for \@citeb:=#2\do {%
    \@ifundefined {b@\@citeb}%
      {\@h@ld\@citea\@tempcntb\m@ne{\bf ?}%
      \@warning {Citation `\@citeb ' on page \thepage \space undefined}}%
      {\@tempcnta\@tempcntb \advance\@tempcnta\@ne%
      \@tempcntb\number\csname b@\@citeb \endcsname \relax%
      \ifnum\@tempcnta=\@tempcntb 
        \ifx\@h@ld\relax%
          \edef \@h@ld{\@citea\csname b@\@citeb\endcsname}%
        \else%
          \edef\@h@ld{\ifmmode{-}\else--\fi\csname b@\@citeb\endcsname}%
        \fi%
      \else
        \@h@ld\@citea\csname b@\@citeb \endcsname%
        \let\@h@ld\relax%
      \fi}%
    \def\@citea{,\penalty\@highpenalty\,}%
  }\@h@ld
}{#1}}

\def\@citeb#1#2{{[#1]\if@tempswa , #2\fi}}
\def\@citeu#1#2{{$^{#1}$\if@tempswa , #2\fi }}
\def\@citep#1#2{{#1\if@tempswa , #2\fi}}

\def\bcites{         
        \catcode`\@=11
        \let\@cite=\@citeb
        \catcode`\@=12
}

\def\upcites{         
        \catcode`\@=11
        \let\@cite=\@citeu
        \catcode`\@=12
}

\def\plaincites{      
        \catcode`\@=11
        \let\@cite=\@citep
        \catcode`\@=12
}

\newcount\hour
\newcount\minute
\newtoks\amorpm
\hour=\time\divide\hour by 60
\minute=\time{\multiply\hour by 60 \global\advance\minute by-\hour}
\edef\standardtime{{\ifnum\hour<12 \global\amorpm={am}%
        \else\global\amorpm={pm}\advance\hour by-12 \fi
        \ifnum\hour=0 \hour=12 \fi
        \number\hour:\ifnum\minute<10 0\fi\number\minute\the\amorpm}}
\edef\militarytime{\number\hour:\ifnum\minute<10 0\fi\number\minute}

\def\draftlabel#1{{\@bsphack\if@filesw {\let\thepage\relax
   \xdef\@gtempa{\write\@auxout{\string
      \newlabel{#1}{{\@currentlabel}{\thepage}}}}}\@gtempa
   \if@nobreak \ifvmode\nobreak\fi\fi\fi\@esphack}
        \gdef\@eqnlabel{#1}}
\def\@eqnlabel{}
\def\@vacuum{}
\def\marginnote#1{}
\def\draftmarginnote#1{\marginpar{\raggedright\scriptsize\tt#1}}
\def\draft{
        \pagestyle{plain}
        \overfullrule=2pt
        \oddsidemargin -.5truein
        \def\@oddhead{\sl \phantom{\today\quad\militarytime} \hfil
        \smash{\Large\sl DRAFT} \hfil \today\quad\militarytime}
        \let\@evenhead\@oddhead
        \let\label=\draftlabel
        \let\marginnote=\draftmarginnote
        \def\ps@empty{\let\@mkboth\@gobbletwo
        \def\@oddfoot{\hfil \smash{\Large\sl DRAFT} \hfil}
        \let\@evenfoot\@oddhead}
        \def\@eqnnum{(\theequation)\rlap{\kern\marginparsep\tt\@eqnlabel}%
        \global\let\@eqnlabel\@vacuum}  }

\begin{document}


\hfill UTHET-02-1001

\vspace{-0.2cm}

\begin{center}
\Large
{\bf Quasinormal modes of large AdS black holes\footnote{Research supported in part by the DoE under grant DE-FG05-91ER40627.}}
\normalsize

\vspace{0.8cm}

{\bf Suphot Musiri\footnote{smusiri@utk.edu} and George Siopsis\footnote{gsiopsis@utk.edu}}\\ Department of Physics
and Astronomy, \\
The University of Tennessee, Knoxville, \\
TN 37996 - 1200, USA.
\end{center}

\vspace{0.8cm}
\large
\centerline{\bf Abstract}
\normalsize
\vspace{.5cm}

We develop a perturbative approach to the solution of the scalar wave equation
for a large AdS black hole. In three dimensions, our method coincides with the
known exact solution. We discuss the five-dimensional case in detail and apply our
procedure to the Heun equation. We calculate the quasi-normal modes analytically
and obtain good agreement with numerical results for the low-lying frequencies.

\newpage


Quasi-normal modes describe small perturbations of a black hole at equilibrium~\cite{bibq1,bibq2,bibq3,bibq4,bibq5,bibq6,bibq7,bibq8,bibq9,bibq10,bibq11,bibq12,bibq13}.
A black hole is a thermodynamical system whose (Hawking) temperature and
entropy are given in terms of its global characteristics (total mass, charge
and angular momentum). In general, quasi-normal modes are obtained by solving
a wave equation for small fluctuations subject to the conditions that the
flux be ingoing at the horizon and outgoing at asymptotic infinity. One then
obtains a discrete spectrum of complex frequencies. The imaginary part of these
frequencies determines the decay time of the small fluctuations. On account
of the AdS/CFT correspondence, the quasi-normal modes for an AdS black hole
are expected to
correspond to perturbations of the dual CFT. The establishment of such a
correspondence is hindered by the difficulties in solving the wave equation.
In three dimensions, the wave equation is a hypergeometric equation and can therefore be solved~\cite{bibq13}. In five dimensions, the wave equation turns into
a Heun equation which is unsolvable~\cite{bibq15}. Numerical results have been obtained
in four, five and seven dimensions~\cite{bibq2,bibq14}.

We discuss an analytic method of solving the wave
equation in the case of
a large black hole living in AdS space. The method is based on a
perturbative expansion
of the wave equation in the dimensionless parameter $\omega / T_H$, where $\omega$
is the frequency of the mode and $T_H$ is the (high) Hawking temperature of the black hole.
Such an expansion is no trivial matter, for the dependence of the wavefunction on $\omega / T_H$ changes as one moves from the asymptotic boundary of AdS space to the horizon
of the black hole.
The zeroth-order approximation is chosen to be an appropriate hypergeometric equation so that higher-order corrections are indeed of higher order in $\omega /T_H$.
In three dimensions, this hypergeometric equation is exact. In five dimensions,
we show that the low-lying quasi-normal modes obtained from the zeroth-order approximation are in good agreement with numerical results~\cite{bibq2,bibq14}.
The first-order correction is also calculated.
Our method is generalizable to higher dimensions where the number of singularities of the wave equation is larger.


The metric of a $d$-dimensional AdS black hole may be written as
\be
ds^2 = \left( \frac{r^2}{R^2} +1 - \frac{\omega_{d-1} M}{r^{d-3}} \right) dt^2 + \frac{dr^2}{ \left( \frac{r^2}{R^2} +1 - \frac{\omega_{d-1} M}{r^{d-3}} \right) } +r^2 d \Omega_{d-2}^2
\ee
where $R$ is the AdS radius and $M$ is the mass of the black hole.
For a large black hole, the metric simplifies to
\be
ds^2 = \left( \frac{r^2}{R^2} - \frac{\omega_{d-1} M}{r^{d-3}} \right) dt^2 + \frac{dr^2}{ \left( \frac{r^2}{R^2} - \frac{\omega_{d-1} M}{r^{d-3}} \right) } +r^2 ds^2 (\mathbb{E}^{d-2})
\ee
The Hawking temperature is
\be
T_H = \frac{d-1}{4\pi}\; \frac{r_h}{R^2}
\ee
where $r_h$ is the radius of the horizon,
\be
r_h = R\; \left[ \frac{\omega_{d-1} M}{R^{d-3}} \right] ^{1/(d-1)}
\ee
The scalar wave equation is
\be\label{eq5}
\frac{1}{\sqrt{g}} \p_A \sqrt{g}g^{AB}\p _B \Phi = m^2 \Phi \ee
We are interested in solving this equation in the massless case ($m=0$) for
a wave which is ingoing at the horizon and vanishes at infinity. These boundary
conditions yield a discrete set of complex frequencies (quasi-normal modes).


We start with a review of the three-dimensional case where the assumption of a
large black hole is redundant and the wave equation may be solved exactly.
Indeed, in three dimensions
$(d=3)$, the metric reads
\be
ds^2 = \frac{1}{R^2}\; \left( r^2 -r_h^2 \right) dt^2 +\frac{R^2\; dr^2}{ \left( r^2 - r_h^2 \right) }+ r^2 dx^2
\ee
independently of the size of the black hole.
The wave equation is
\be
\frac{1}{R^2\; r}\p_r \left( r^3 \left( 1- \frac{r_h^2}{r^2}\right) \p_r \Phi\right) -\frac{R^2}{r^2 - r_h^2 }\p_t^2 \Phi + \frac{1}{r^2}\p_x^2 \Phi -m^2\Phi =0 \nonumber
\ee
The solution may be written as
\be
\Phi = e^{i(\omega t-px) }\Psi (y) ,\ \ \ \ \ y = \frac{r_h^2}{r^2}
\ee
where $\Psi$ satisfies
\be
y^2 (y-1)\left( (y-1) \Psi' \right)' +\hat\omega^2\, y\Psi +\hat p^2\, y(y-1)\Psi +\four\hat m^2\, (y-1)\Psi = 0
\ee
in the interval $0<y<1$, and we have introduced the dimensionless variables
\be
\hat\omega = \frac{\omega R^2}{2r_h} = \frac{\omega}{4\pi T_H},\ \ \ \hat p^2 = \frac{pR}{2r_h} = \frac{p}{4\pi R T_H}\;,\quad
\hat m = mR
\ee
Two independent solutions are obtained by examining the behavior near the horizon
($y\to 1$),
\be\label{eq11} \Psi_\pm \sim (1-y)^{\pm i\hat\omega}\ee
where $\Psi_+$ is outgoing and $\Psi_-$ is ingoing. A different set of linearly
independent solutions is obtained by studying the behavior at large $r$ ($y\to 0$).
We obtain
\be \Psi\sim y^{h_\pm}
\quad,\quad
h_\pm = \frac{1}{2} \pm \frac{1}{2}\sqrt{1 +\hat m^2}
\ee
In the massless case ($m = 0$), we have $h_+=1$ and $h_-=0$, so one of the
solutions contains logarithms. For quasi-normal modes, we are interested in the
analytic solution. This may be written in terms of a Hypergeometric function as
\be \Psi (y) = y(1-y)^{i\hat\omega} {}_2F_1 (1+i(\hat\omega + \hat p), 1+i(\hat\omega - \hat p); 2; y)\ee
Near the horizon, this solution is a mixture of ingoing and outgoing waves. Using
standard identities involving Hypergeometric functions, we obtain
\be \Psi \sim A (1-y)^{-i\hat\omega} + B(1-y)^{i\hat\omega}\ee
where
\be\label{eq15} A=
\frac{\Gamma(2i\hat\omega)}{\Gamma(1+i(\hat\omega + \hat p))\Gamma(1+i(\hat\omega - \hat p))}
\quad,\quad B= \frac{\Gamma(-2i\hat\omega)}{\Gamma(1-i(\hat\omega + \hat p))\Gamma(1-i(\hat\omega - \hat p))}\ee
as $y\to 1$. Since $\Psi$ must be a linear combination of $\Psi_+$ and $\Psi_-$,
we deduce
\be \Psi = A\Psi_- + B\Psi_+\ee
on account of eq.~(\ref{eq11}).
For quasi-normal modes, we demand that $\Psi$ be purely ingoing at the horizon, so
we need to set
\be B=0\ee
with $B$ given in~(\ref{eq15}).
The solutions to this equation are the quasi-normal frequencies. Explicitly,
\be \hat\omega = \pm \hat p  -in\quad,\quad n=1,2,\dots\ee
a discrete set of complex frequencies with negative imaginary part, as expected~\cite{bibq2}. Notice that we obtained two sets of frequencies, with opposite real parts.


In five dimensions ($d=5$), the wave equation~(\ref{eq5}) reads
\be
\frac{1}{r^3}\p_r r^5\left( 1- \frac{r_h^4}{r^4} \right) \p_r \Phi -\frac{R^4}{ r^2\; \left( 1- \frac{r_h^4}{r^4} \right) }\p_{t}^2\Phi - \frac{R^2}{r^2}\; \vec\nabla^2\Phi - m^2R^2\Phi = 0
\ee
Let 
\be 
\Phi = e^{i(\omega t - \vec p\cdot \vec x)} \Psi (r)
\ee
and change parameter $r$ to $y$,
\be
y = \frac{r_h^2}{r^2} 
\ee
The wave equation then becomes
\be
y^3 (1-y^2)\left( \frac{1}{y}(1-y^2) \Psi' \right)' + \frac{\hat\omega^2}{4}\, y\Psi + \frac{\hat p^2}{4}\, y(1-y^2)\Psi - \four \hat m^2 (1-y^2)\Psi =0
\ee
where
\be
\hat\omega = \frac{\omega R^2}{r_h} = \frac{\omega}{\pi T_H}, \ \ \ \ \
\hat p = \frac{|\vec p|R}{r_h} = \frac{|\vec p|}{\pi R T_H}, \ \ \ \ \
\hat m = m R
\ee
Near the horizon we obtain in- and out-going waves
\be \Psi_\pm \sim (1-y)^{\pm i\hat\omega/4}\ee
At large $r$ ($y\to 1$) there is only one analytic solution for $m=0$ (the other
one contains logarithms). It may be written as
\be\label{eq25}
\Psi (y) = y^2(1-y)^{-i\hat\omega/4} \left(\frac{1+y}{2} \right)^{-\hat\omega/4} F(y)
\ee
where we isolated the singularities at $y=0, \pm 1$. We divided $(1+y)$ by
$2$ in order not to obtain a contribution from this factor at the horizon
($y\to 1$). This introduces an irrelevant constant, but one ought to be careful
with $\hat\omega$ dependent constants in perturbation theory (expansion in $\hat\omega$). It should also be pointed out that we could have chosen the exponent $+\hat\omega/4$ instead of $-\hat\omega/4$ at the $y=-1$ singularity (eq.~(\ref{eq25})). This does not alter the
quasi-normal modes, but in the perturbative expansion, each choice only produces
half the modes. The two sets of modes have the same imaginary parts, but opposite
real parts, as we shall see.

$F$ is the Heun function satisfying the equation~\cite{bibq15}
\be
F'' + \left( \frac{3}{y} +\frac{1-i\hat\omega/2}{y-1} +\frac{1-\hat\omega/2}{y+1} \right) F'
+ \frac{(2-(1+i)\hat\omega/4)^2 y -q}{y(y^2-1)}\; F =0 \label{w1}
\ee
where
\be
q = \frac{3(-1+i)}{4}\,\hat\omega -\frac{\hat p^2}{4} +\frac{\hat\omega^2}{4}
\ee
To find the behavior of $F$ near the horizon, let us write the Heun equation as
\be (\mathcal{H}_0 +\mathcal{H}_1) F = 0\ee
where
\bea \mathcal{H}_0 &=& x(1-x)\frac{d^2}{dx^2} + (2-{\textstyle{\frac{1-i}{4}}}\,\hat\omega-(3-{\textstyle{\frac{1+i}{4}}}\,\hat\omega)x)\frac{d}{dx} -
\frac{1}{4} ((2-{\textstyle{\frac{1+i}{4}}}\,\hat\omega)^2-q)\nonumber\\
\mathcal{H}_1 &=& (1-\sqrt x) \left({\textstyle{\frac{1-i}{4}}}\,\hat\omega\,\frac{d}{dx} + \frac{q}{4\sqrt x} \right)\eea
and we changed variables to $x=y^2$. We shall treat $\mathcal{H}_1$ as a perturbation. It has been chosen so that the perturbation series is an expansion in
$\hat\omega \sim \omega / T_H$. Expanding the wavefunction,
\be F = F_0 + F_1 + \dots\ee
the zeroth-order equation
\be \mathcal{H}_0 F_0 = 0\ee
has analytic solution the Hypergeometric function
\be\label{eq32} F_0(x) = {}_2F_1 ( 1+(-{\textstyle{\frac{1+i}{4}}}\,\hat\omega+\sqrt q)/2, 1+(-{\textstyle{\frac{1+i}{4}}}\,\hat\omega-\sqrt q)/2; 2-{\textstyle{\frac{1-i}{4}}}\,\hat\omega; x)\ee
Its behavior at the horizon ($x\to 1$) is
\be F_0(x) \sim A_0 + B_0 (1-x)^{i\hat\omega/2}\ee
where
\bea A_0 &=& \frac{\Gamma (2-{\textstyle{\frac{1-i}{4}}}\,\hat\omega)\Gamma(i\hat\omega/2)}{\Gamma (1-({\textstyle{\frac{1-3i}{4}}}\,\hat\omega +\sqrt q)/2)
\Gamma (1-({\textstyle{\frac{1-3i}{4}}}\,\hat\omega -\sqrt q)/2)}\nonumber\\
B_0 &=& \frac{\Gamma (2-{\textstyle{\frac{1-i}{4}}}\,\hat\omega)\Gamma(-i\hat\omega/2)}{\Gamma (1-({\textstyle{\frac{1+i}{4}}}\,\hat\omega +\sqrt q)/2)
\Gamma (1-({\textstyle{\frac{1+i}{4}}}\,\hat\omega -\sqrt q)/2)}\eea
An approximation to the quasi-normal modes is obtained by setting
\be\label{eq1} B_0 = 0\ee
The solutions to this equation are
\be {\textstyle{\frac{1+i}{4}}}\,\hat\omega \pm\sqrt q = 2n\,, \ \ \ \ n= 1,2,\dots\ee
For $\hat p = 0$, they are shown on Table 1, column {\em (a)}. They all have negative real parts. The set of modes with positive real parts
may be obtained by replacing the factor $(\frac{1+y}{2})^{-\hat\omega /4}$
with $(\frac{1+y}{2})^{+\hat\omega /4}$ in eq.~(\ref{eq25}) (capturing the alternative behavior of the
wavefunction near the singularity $y=-1$). The low-lying frequencies
are close to the exact values obtained by numerical methods (Table 1,
column {\em (c)}~\cite{bibq14}).
For first-order perturbation theory, we need the expansion of $B_0$ in $\hat\omega$,
\be\label{eq37} B_0 = \frac{1}{-i\hat\omega /2} \; \left\{ 1 + \left( -1 + \frac{\pi^2}{8} \right) {\textstyle{\frac{1-i}{4}}}\,\hat\omega
+ \dots \right\}\ee
To calculate the first-order correction, we also need
another linearly independent solution of the zeroth-order wave equation.
We shall use
\be\label{eq38} G_0(x) = F_0(x)\int^x \frac{W_0(x')\, dx'}{(F_0(x'))^2}\ee
where $W_0$ is the Wronskian
\be W_0 \equiv F_0G_0' - G_0F_0' = x^{-2+(1-i)\hat\omega/4} (1-x)^{-1+i\hat\omega/2}\ee
The equation satisfied by the first-order contribution to the wavefunction $F$ is
\be \mathcal{H}_0 F_1 = -\mathcal{H}_1 F_0\ee
The solution (satisfying $F_1(0) = 0$) may be written as
\be\label{eq41} F_1(x) =
G_0(x)\int_0^x \frac{F_0(x')\mathcal{H}_1 F_0(x')\, dx'}{W_0(x')}
- F_0(x)\int_0^x \frac{G_0(x')\mathcal{H}_1 F_0(x')\, dx'}{W_0(x')}\ee
Expanding in $\hat\omega$, we obtain from eqs.~(\ref{eq32}) and (\ref{eq38}),
respectively,
\be F_0(x) = \frac{1-(1-x)^{i\hat\omega/2}}{i\hat\omega\, x/2} + \dots\quad,\quad G_0(x) = -\frac{1}{x} + \dots \ee
The first term in $F_1$ (eq.~(\ref{eq41})) does not contribute at lowest order. After some algebra, we arrive at
\be F_1(x) = C\, {\textstyle{\frac{1-i}{4}}}\,\hat\omega\, F_0(x) + \dots\ee
where
\bea C &=& \int_0^1 dx\; \left\{ -\frac{1}{x(1+\sqrt x)} - \frac{1-\sqrt x}{x^2}
\left( 1 + \frac{3\sqrt x}{4} \right) \ln (1-x) \right\}\nonumber\\
&=& 1+\ln 2-\frac{\pi^2}{8} \label{eq44}\eea
Notice that the divergences from the contributing terms in the integral in eq.~(\ref{eq44}) at
both ends ($x=0, 1$) all cancel each other, resulting in a finite expression for
$C$, as required for the validity of our perturbative expansion.
To first order, the behavior at the horizon is
\be F(x) \approx F_0(x) + F_1 (x) \sim A_1 + B_1 (1-x)^{i\hat\omega/2}\ee
Using eqs.~(\ref{eq37}) and (\ref{eq44}), we obtain
\be\label{eq46} B_1 = B_0 + {\textstyle{\frac{1-i}{2}}}\, C\, = \frac{1}{-i\hat\omega /2}\; \left\{ 1 + \ln 2 {\textstyle{\frac{1-i}{4}}}\,\hat\omega
+ \dots \right\}\ee
At this order, we obtain an approximation to the lowest quasi-normal frequency
by setting $B_1 = 0$ in eq.~(\ref{eq46}). We find
\be\label{eq2} \hat\omega_1 = -\frac{2(1+i)}{\ln 2} = -2.89 (1+i)\ee
which is a good approximation to the exact result $\hat\omega_1 = \pm 3.12 -
2.75 i$ obtained by numerical techniques~\cite{bibq2,bibq14}. (The mode with
positive real part is obtained by replacing the exponent $-\hat\omega /4$
by $+\hat\omega/4$ at the singularity $y=-1$ (eq.~(\ref{eq25})), as already explained). Higher-order modes may be obtained by calculating higher-order perturbative corrections.
At the $n$th perturbative order we obtain an $n$th order polynomial in $\hat\omega$ whose roots
are an approximation to the $n$ lowest quasi-normal frequencies.

Our method may
be straightforwardly extended to higher dimensions even though the wave equation
possesses more (complex) singularities than the Heun equation in five dimensions. It would be interesting to investigate the implications
of our perturbative approach to the AdS/CFT correspondence.
\begin{table}
\begin{center}
\begin{tabular}{cccc}
\hline
$n$ & {\em (a)} & {\em (b)} & {\em (c)} \\ \hline
1 & $\pm 1.19-3.03i$ & $\pm 2.89-2.89i$ & $\pm 3.12-2.75i$ \\
2 & $\pm 5.01-7.65i$  & & $\pm 5.17-4.76i$  \\
3 & $\pm 9.01-11.78i$ & & $\pm 7.19-6.77i$  \\
4 & $\pm 13.01-15.84i$ & & $\pm 9.20-8.77i$  \\
5 & $\pm 17.00-19.87i$ & & $\pm 11.20-10.77i$  \\
6 & $\pm 21.00-23.90i$ & & $\pm 13.21-12.78i$  \\
7 & $\pm 25.00-27.92i$ & & $\pm 15.21-14.78i$  \\
8 & $\pm 29.00-31.93i$ & & $\pm 17.21-16.78i$  \\
9 & $\pm 33.00-35.93i$ & & $\pm 19.21-18.78i$  \\
10 & $\pm 37.00-39.94i$ & & $\pm 21.21-20.78i$ \\ \hline
\end{tabular}
\end{center}
\caption{Quasi-normal frequencies in $d=5$: {\em (a)}~zeroth-order untruncated approximation
(eq.~(\ref{eq1})), {\em (b)}~first-order approximation (eq.~(\ref{eq2})),
{\em (c)} numerical results~\cite{bibq14}.}
\end{table}

\end{document}